\documentclass[preprint,showpacs,preprintnumbers,amsmath,amssymb,aps]{revtex4}

\usepackage{graphicx}
\usepackage{dcolumn}
\usepackage{bm}
\usepackage{natbib}
\usepackage{braket}
\usepackage{units}
\usepackage{subfigure}
\usepackage{color}

\renewcommand{\d}{ \text{d}}
\def\lsim{\mathrel{\vcenter{\hbox{$<$}\nointerlineskip\hbox{$\sim$}}}}
\def\gsim{\mathrel{\vcenter{\hbox{$>$}\nointerlineskip\hbox{$\sim$}}}}

\begin{document}

\preprint{DO-TH-12/26}

\title{Pion production by neutrinos in the delta resonance region and possible application to CP searches}

\author{E. A. Paschos}
 \email{paschos@physik.uni-dortmund.de}
\author{Dario Schalla}
 \email{dario.schalla@tu-dortmund.de}
\affiliation{Department of Physics, TU Dortmund, D-44221 Dortmund, Germany}

\date{November 6, 2012}

\begin{abstract}
We present the results of extensive calculations for charged and neutral current reactions of neutrinos and antineutrinos in the small $Q^2\lsim\unit[0.20]{GeV^2}$ region. The results include single $\frac{\d \sigma}{\d Q^2}$ and double $\frac{\d \sigma^{(A)} }{\d Q^2 \d E_\pi}$ differential cross sections at energies relevant for oscillation experiments. We include nuclear corrections in the Adler-Nussinov-Paschos model and point out that on isoscalar nuclear targets there are charge symmetry relations that hold in extended kinematic regions. We discuss how the results can be used in long baseline experiments in order to study oscillation parameters and search for CP asymmetries.
\end{abstract}

\pacs{13.15.+g, 13.60.Le, 14.20.Gk, 25.30.Pt, 21.65.Jk, 11.30.Er}

\maketitle

\section{Introduction}
\label{sec:Introduction}
Neutrino interactions have been studied in the resonance region in terms of form factors and have been compared with available data~\cite{Rein198179,Paschos:2003qr,Lalakulich:2005cs,Hernandez:2007qq,Lalakulich:2010ss,SajjadAthar:2008hi,Graczyk:2009qm,Sato:2003rq} with considerable success. After the early experiments with larger errors, there is a new generation of experiments that provide more accurate data. In addition, the oscillation phenomena and the finite mixing angle $\theta_{13}$~\cite{An:2012eh,Ahn:2012nd} require more accurate theoretical predictions in order to decipher properties of oscillations including CP asymmetries. All these require on the theoretical side estimates of amplitudes that are reliable.

One property of production cross sections is the fact that for $m_N\nu\gg Q^2$ and $Q^2 \approx \mathcal{O} (m_\pi^2)$ the axial contribution is given by the partially conserved axialvector current (PCAC). We worked out this framework for resonance production~\cite{Paschos:2011ye} and for coherent scattering on nuclei it has been reported by two groups~\cite{Paschos:2005km,Paschos:2009ag,Berger:2008xs}. In this article we wish  to present our results for many reactions, calculate explicitly the energy spectra of the produced pions and point out special properties. For example, in the production of the delta resonance in the energy region $E_\nu = 1.0$ to $\unit[2.0]{GeV}$ the vector squared and the interference contributions are almost equal~\cite{Paschos:2011ye}. They add up for neutrinos and cancel each other for antineutrinos leaving the axial current squared as the dominant contribution in the latter reaction. This will be tested in the experiments. It will also be useful for measuring the flux by using this and additional channels in the nearby detector and to predict the yield in the far away detector in order to establish deviations from normal oscillations, like the presence of a CP violating phase. The above property for antineutrino reactions is especially useful because the flux of the antineutrino beam will be smaller.
For antineutrino interactions we find the cross section on a proton target to be smaller than on a neutron. Early qualitative evidence for this ordering is already available for integrated cross sections~\cite{Bolognese:1978yz,Grabosch:1988gw}.

In this article we adopt the neutrino reaction
\begin{align}
\nu_\mu + p \rightarrow \mu^- + \Delta^{++}
\label{eq:vCCp}
\end{align}
as benchmark for comparing other reactions of charged and neutral currents. For $W\leq\unit[1.6]{GeV}$ the delta resonance dominates and it has been shown that there is only a small background of 10\% for other reactions. We shall assume that the $I=\nicefrac{3}{2}$ amplitude dominates and thus obtain the antineutrino and neutral current reactions. Later on, we may revisit the topic in order to study the changes brought about when we introduce a small nonresonant background.

In section~\ref{sec:EnergySpectrumAndAngularDependenceOfPions} we calculate the energy and angular spectrum of the pions for the small region of $Q^2$. For this we introduce the $\pi p \rightarrow \Delta \rightarrow \pi p$ cross section in the rest frame of the resonance and then transform it to the laboratory frame of the neutrino interaction. Details of the calculation are given in section~\ref{sec:EnergySpectrumAndAngularDependenceOfPions} and in the appendix.

One worrisome aspect concerns the modifications brought about by the subsequent rescattering of pions in nuclei. Here we make the observation that charge symmetry predicts similar corrections for $\pi^+$ and $\pi^-$ interactions on isoscalar nuclei. We shall use this property.
For nuclear corrections we adopt a model for rescatterings which is both simple and transparent~\cite{Adler:1974qu}. In fact, since some of the neutrino-nucleon reactions are predicted, we can use them to test the accuracy of nuclear corrections.

For isoscalar targets many structure functions are related by charge symmetry at both steps of the reaction: the initial neutrino-nucleon scattering and the subsequent pion nucleus interactions. Consequently, several relations follow to be tested experimentally. Finally, the results for $Q^2 \leq \unit[0.20]{GeV^2}$ are useful for matching them to the form factors at higher values of $Q^2$.

The contents of the article are arranged as follows. In section~\ref{sec:ChargedAndNeutralCurrentReactions} we
define the general method and present charged and neutral current
cross sections. We point out several regularities that are inherent in
the cross sections.  Results for pion spectra in energy $E_\pi$ from the axial current are presented in section~\ref{sec:EnergySpectrumAndAngularDependenceOfPions}. For these
calculations we introduce the energy and angular dependence of the
reaction $\pi + p \rightarrow \textmd{resonance} \rightarrow \pi + p$ in the rest frame of the
resonance and transform it to the laboratory frame. The details of the
Lorentz transformation are given in the appendix. In section~\ref{sec:NuclearCorrections} we
compute nuclear corrections for isoscalar targets
and pay special attention to find charge symmetric reactions. The
results will be useful for long baseline experiments in which we
point out that in our kinematic region products of cross sections $\otimes$ nuclear corrections will be
very similar for the regenerated $\nu_e$ and $\overline{\nu}_e$ beams. Since the mass
$m_e$ of the electron is very small we repeated the calculations with $m_e$
and the results are shown in figure~\ref{fig:electron}. The article closes with a
general discussion.

The method can be extended to higher resonances in order to fill the transition region between resonances and deep inelastic scattering. In fact, another group~\cite{Kamano:2012he} uses a dynamic hadron model for this purpose.

\section{Charged and neutral current reactions}
\label{sec:ChargedAndNeutralCurrentReactions}
\begin{figure}
\includegraphics{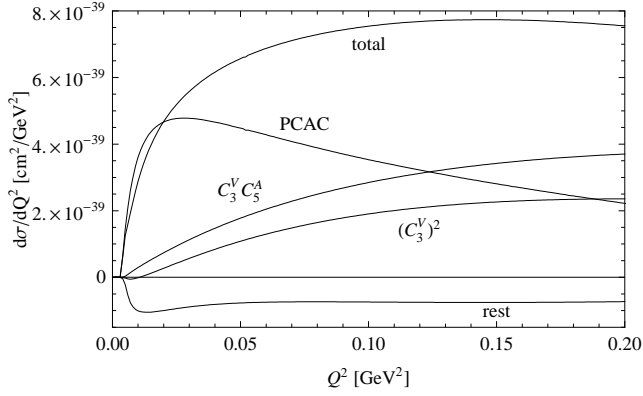}
\caption{Various contributions to the differential cross section at $E_\nu = \unit[1]{GeV}$ (from reference~\cite{Paschos:2011ye}).}
\label{fig:PCACold}
\end{figure}

We consider the reactions listed in table~\ref{tab:ISOreactions}. We take reaction~(\ref{eq:vCCp}) as a standard where for $X^{++}=\Delta^{++}$ there is only the $A^{\nicefrac{3}{2}}$ amplitude. The amplitude for
\begin{align}
\nu_\mu + n \rightarrow \mu^- + \Delta^+
\label{eq:vCCn}\end{align}
is given by the Clebsch-Gordan coefficient (CGC) shown in the fourth column of table~\ref{tab:ISOreactions}. The CGC is the same for vector and axial-vector form factors and reduces the cross section by a factor of $\nicefrac{1}{3}$. When the subsequent decays to specific pion-nucleus channels are considered, there are additional CGCs. The results for vector, axial-vector and interference terms were computed in an earlier article~\cite{Paschos:2011ye} and the results are shown again in figure~\ref{fig:PCACold}. Considering next the reaction
\begin{align}
\overline{\nu}_\mu + n \rightarrow \mu^+ + \Delta^-
\label{eq:avCC}
\end{align}
we use the same structure functions as for~(\ref{eq:vCCp}), since they are related by charge symmetry, but the sign of the interference term in the cross section changes. We note in figure~\ref{fig:PCACold} that the vector and interference terms are approximately equal and cancel each other for antineutrinos, leaving the axial contribution alone.

\begin{table}
\begin{tabular}{|c|c|c|c|c|c|}
\hline 
reaction		& sign of $\mathcal{W}_3$	& lepton mass	& CGC				& $C_i^A\times$	& $C_i^V\times$\\
\hline
$\nu_\mu p \rightarrow \mu^- X^{++}$					& $+$	& $m_\mu$	& 1				& 1	& 1 \\
$\overline{\nu}_\mu p \rightarrow \mu^+ X^0$				& $-$	& $m_\mu$	& $\nicefrac{1}{\sqrt{3}}$	& 1	& 1 \\
$\nu_\mu n \rightarrow \mu^- X^{+}$					& $+$	& $m_\mu$	& $\nicefrac{1}{\sqrt{3}}$	& 1	& 1 \\
$\overline{\nu}_\mu n \rightarrow \mu^+ X^-$				& $-$	& $m_\mu$	& 1				& 1	& 1 \\
$\nu_\mu p \rightarrow \nu_\mu X^+$					& $+$	& $0$		& $\nicefrac{1}{\sqrt{3}}$	& y	& x \\
$\overline{\nu}_\mu p \rightarrow \overline{\nu}_\mu X^+$		& $-$	& $0$		& $\nicefrac{1}{\sqrt{3}}$	& y	& x \\
$\nu_\mu n \rightarrow \nu_\mu X^0$					& $+$	& $0$		& $\nicefrac{1}{\sqrt{3}}$	& y	& x \\
$\overline{\nu}_\mu n \rightarrow \overline{\nu}_\mu X^0$		& $-$	& $0$		& $\nicefrac{1}{\sqrt{3}}$	& y	& x \\
\hline
\end{tabular}
\caption{Input quantities and isospin factors for various reactions.}
\label{tab:ISOreactions}

\end{table}

\begin{figure}
\subfigure[$\nu_\mu p \rightarrow \mu^- X^{++}$]{\label{fig:vCCp}\includegraphics[width=0.49\textwidth]{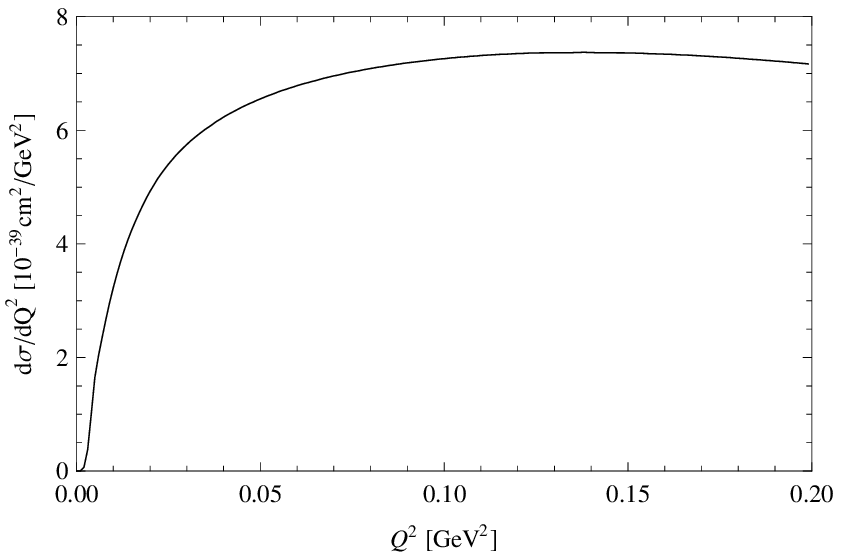}}\hfill
\subfigure[$\overline{\nu}_\mu p \rightarrow \mu^+ X^0$]{\label{fig:avCCp}\includegraphics[width=0.49\textwidth]{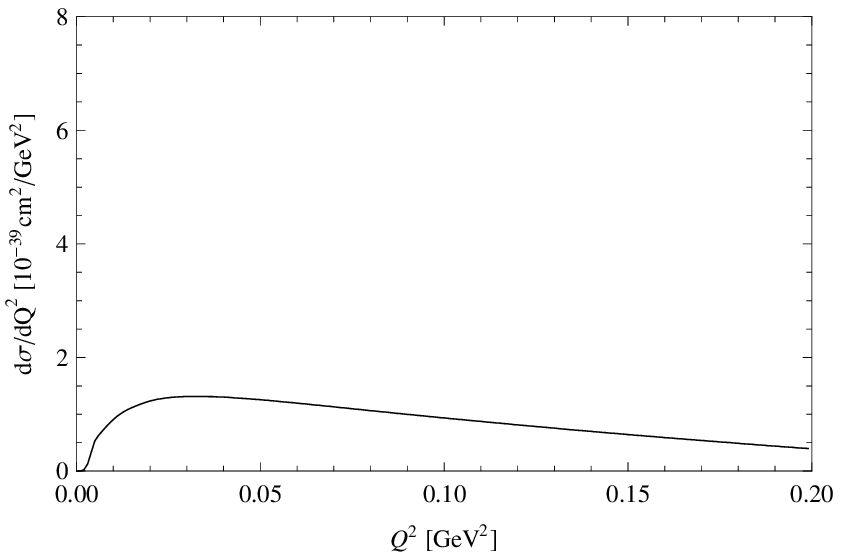}} \\
\subfigure[$\nu_\mu n \rightarrow \mu^- X^+$]{\label{fig:vCCn}\includegraphics[width=0.49\textwidth]{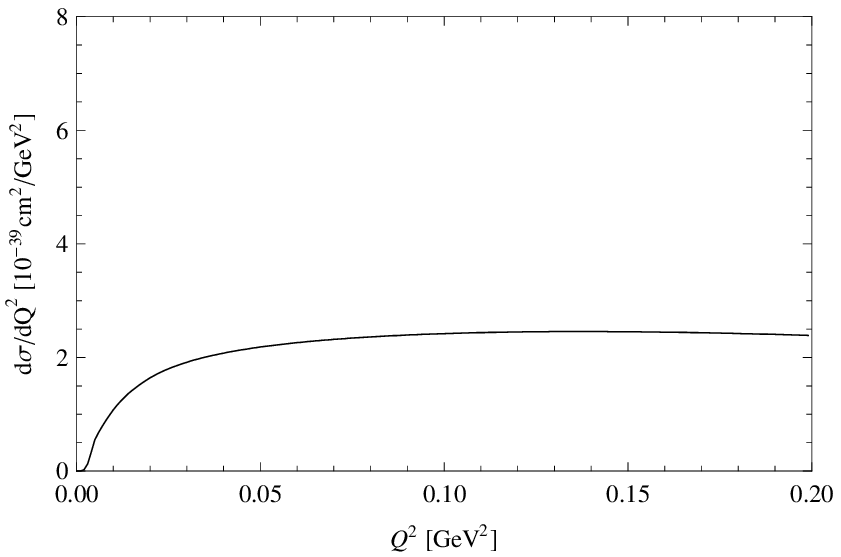}}\hfill
\subfigure[$\overline{\nu}_\mu n \rightarrow \mu^+ X^-$]{\label{fig:avCCn}\includegraphics[width=0.49\textwidth]{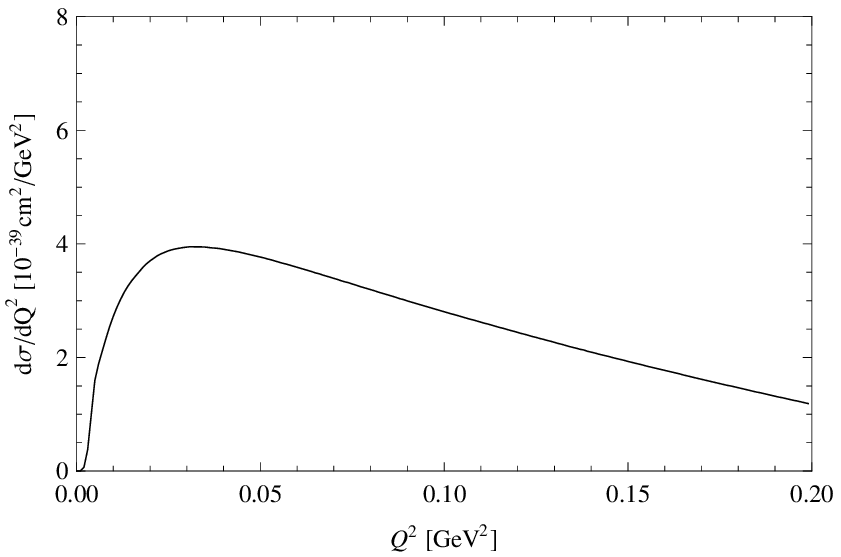}} \\
\caption{Charged current differential cross sections for $E_\nu = \unit[1]{GeV}$.}
\label{fig:IsospinRelationsCC}
\end{figure}

In figure~\ref{fig:IsospinRelationsCC} we show all charged current reactions on proton and neutron targets. 
The cross section for the reaction in equation~(\ref{eq:avCC}) is shown in figure~\ref{fig:avCCn} with its dominant contribution coming from the axial current.
Figures~\ref{fig:vCCn} and~\ref{fig:avCCp} are obtained from~\ref{fig:vCCp} and~\ref{fig:avCCn}, respectively, by applying CGCs of table~\ref{tab:ISOreactions}. In all charged current reactions we kept the muon mass which is responsible for the turning over of the cross section as $Q^2\rightarrow 0$. All these reactions and especially~\ref{fig:avCCn} can be used to confirm and also decipher features of the reactions.

For neutral current reactions there are more changes. The effective interaction is
\begin{align}
\mathcal{H}_{\textmd{eff}} = \frac{G_F^2}{\sqrt{2}} \overline{\nu} \gamma^\mu \left( 1 - \gamma_5 \right) \nu \left[ x \mathcal{V}_\mu^3 + y \mathcal{A}^3_\mu + \gamma \mathcal{V}_\mu^0 \right]
\end{align}
with $\mathcal{V}_\mu^3$ and $\mathcal{A}^3_\mu$ the isovector and $\mathcal{V}_\mu^0$ the isoscalar hadronic currents. The parameters in the hadronic current are given in terms of the weak angle $\theta_W$
\begin{align}
x = 1 - 2 \sin^2\theta_W, \hspace{6mm} y = -1 \hspace{6mm} \textmd{and} \hspace{6mm} \gamma = - \frac{2}{3} \sin^2\theta_W
\end{align}
with $\sin^2\theta_W \approx 0.25$. The value of $y=-1$ gives a constructive $\mathcal{W}_3$ interference term (because of the structure of the lepton current $\overline{\nu} \gamma^\mu \left( 1 - \gamma_5 \right) \nu$), making the neutrino reaction larger than the antineutrino. 
Beyond these parameters there is an overall normalization factor in the amplitudes. In the charged current interaction appears the current
\begin{align}
 \mathcal{A}_\mu^1 + i \mathcal{A}_\mu^2 = \sqrt{2} \left( \frac{\mathcal{A}_\mu^1 + i \mathcal{A}_\mu^2}{\sqrt{2}} \right) = \sqrt{2} \mathcal{A}_\mu^+
\end{align}
and for the neutral current $\mathcal{A}_\mu^3$. The CGCs are valid for the triplet $\left(\mathcal{A}_\mu^+, \mathcal{A}_\mu^3, \mathcal{A}_\mu^- \right)$. Since we have taken the amplitude for reaction~(\ref{eq:vCCp}) as the standard amplitude, we divided the neutral current CGCs in the fourth column of table~\ref{tab:ISOreactions} by $\sqrt{2}$.

\begin{figure}
\subfigure[$\nu_\mu p (n) \rightarrow \nu_\mu X^+ (X^0)$]{\label{fig:vNC}\includegraphics[width=0.49\textwidth]{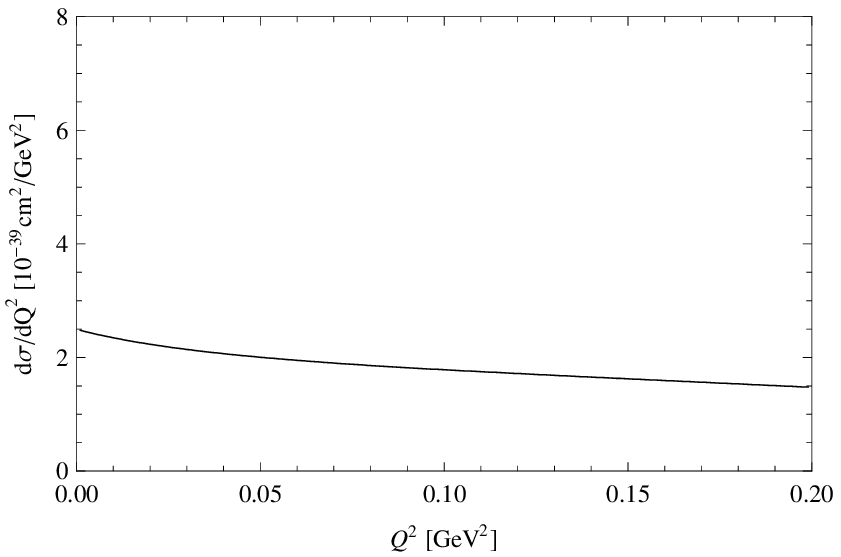}}\hfill
\subfigure[$\overline{\nu}_\mu p (n) \rightarrow \overline{\nu}_\mu X^+ (X^0)$]{\label{fig:avNC}\includegraphics[width=0.49\textwidth]{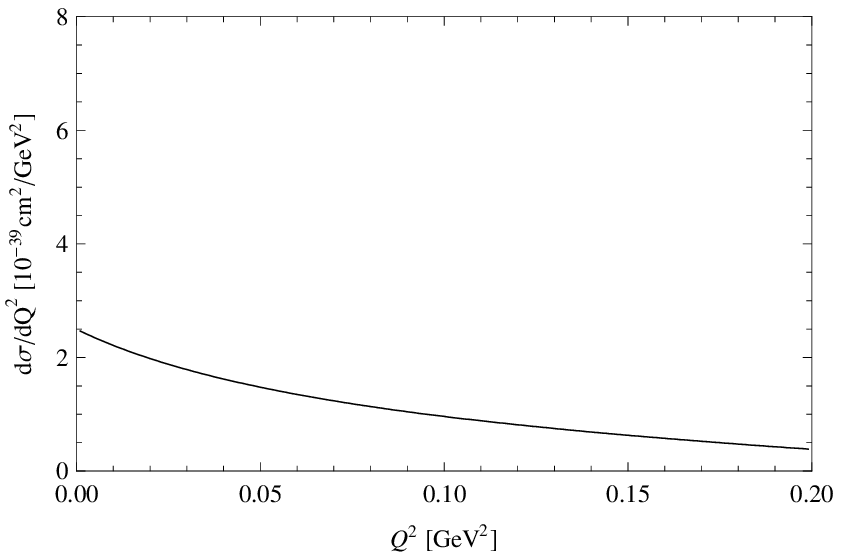}} 
\caption{Differential cross sections for neutral current and $E_\nu = \unit[1]{GeV}$.}
\label{fig:IsospinRelationsNC}
\end{figure}

An additional property of neutral current reactions is
\begin{align}
\sigma (\nu p \rightarrow \nu \Delta^+ ) & = \sigma (\nu n \rightarrow \nu \Delta^0 ) \\
\sigma (\overline{\nu} p \rightarrow \overline{\nu} \Delta^+ ) & = \sigma (\overline{\nu} n \rightarrow \overline{\nu} \Delta^0 ).
\end{align}
which follows from charge symmetry. The calculated differential cross sections for neutral currents are shown in figure~\ref{fig:IsospinRelationsNC}. The zero mass of the neutrino assures nonzero values for the cross sections at $Q^2=0$ which is the exact point from PCAC where neutrino and antineutrino cross sections are equal.

\section{Energy spectrum and angular dependence of pions}
\label{sec:EnergySpectrumAndAngularDependenceOfPions}
Several other variables are measured or have been reported~\cite{AguilarArevalo:2010bm,AguilarArevalo:2010xt,AguilarArevalo:2009ww,Osta,Snider} in recent experiments. Among them are the energy and angle of the pion in the laboratory frame. In our approach the lepton part of the interaction factorizes from the pion-target interaction. For this reason the pion-proton interaction can be computed in the rest-frame of the resonance and then transformed to the laboratory frame. For the delta resonance the angular and energy dependence is known~\cite{Gasiorowicz}:
\begin{align}
\frac{\d \sigma}{\d \Omega_{\text{cm}}} = \sigma_{\pi^+p}(W) \frac{1+ 3 \cos^2 \theta_\text{cm}}{8\pi}
\label{eq:sigmapip}\end{align}
with $\sigma_{\pi^+p}(W)$ given by experimental data~\cite{Nakamura:2010zzi}. As mentioned already, we are interested in the production in the rest frame of the proton, which is obtained by a Lorentz transformation. We present the kinematics and the transformation in the appendix.

Folding this formula with the remaining weak vertex, we derive a triple differential cross section
\begin{align}
\frac{\d \sigma^{(A)} \left( E_\nu, Q^2, W, E_\pi^{\textmd{lab}} \right)}{\d Q^2 \d W \d E_\pi^{\textmd{lab}}} = \frac{G_F^2 |V_{ud}|^2}{8\pi^2} \frac{W}{m_N} \left( \frac{\nu}{E_\nu^2} \frac{\tilde{L}_{00}}{Q^2} f_\pi^2 \right) \frac{\d \sigma_{\pi^+p}(W)}{\d E_\pi^{\textmd{lab}}}
.\label{eq:sigmaA}\end{align}
The factor before the $\pi^+p$ scattering depends on variables $E_\nu, Q^2, \nu$ which is typical for neutrino interactions. The function $\tilde{L}_{00}$ defines the square of the weak vertex contributing to the helicity zero polarization and includes the mass of the muon. It has been given in the articles~\cite{Paschos:2005km,Paschos:2009ag}. We emphasize that formula~(\ref{eq:sigmaA}) gives the contribution of the axial current alone, which for the antineutrino reaction it is a good approximation, especially in the low $Q^2$ region. For the neutrino induced reaction the vector and interference terms must be included as was done in~\cite{Paschos:2008gs}, which increases the cross section.

One obtains $\frac{\d \sigma_{\pi^+p}(W)}{\d E_\pi^{\textmd{lab}}}$ from equations~(\ref{eq:apa}) to~(\ref{eq:apc}) in the appendix by introducing the Jacobian which transforms the cross section from the rest frame of the resonance to the laboratory frame.

\begin{figure}
\subfigure[$E_\nu = 1$ GeV]{\label{fig:dsdQsqdEpilabofEpilab1}\includegraphics[width=0.49\textwidth]{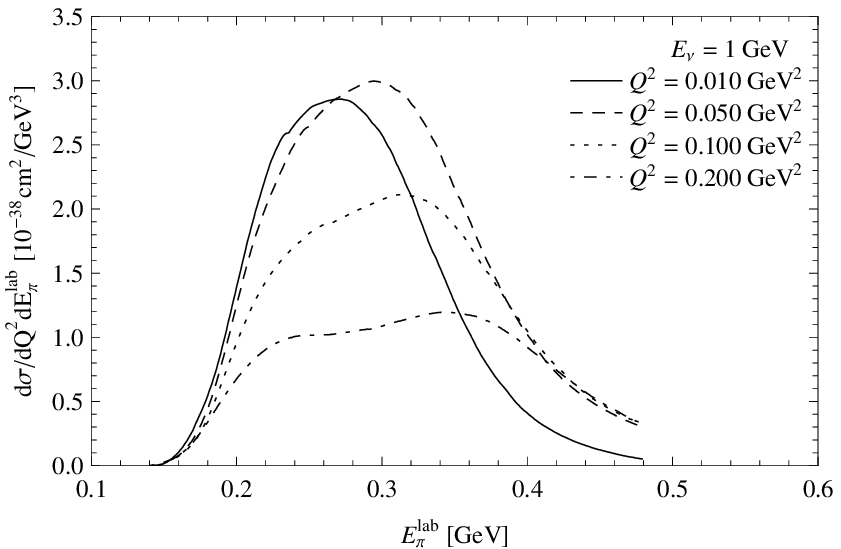}}\hfill
\subfigure[$E_\nu = 5$ GeV]{\label{fig:dsdQsqdEpilabofEpilab5}\includegraphics[width=0.49\textwidth]{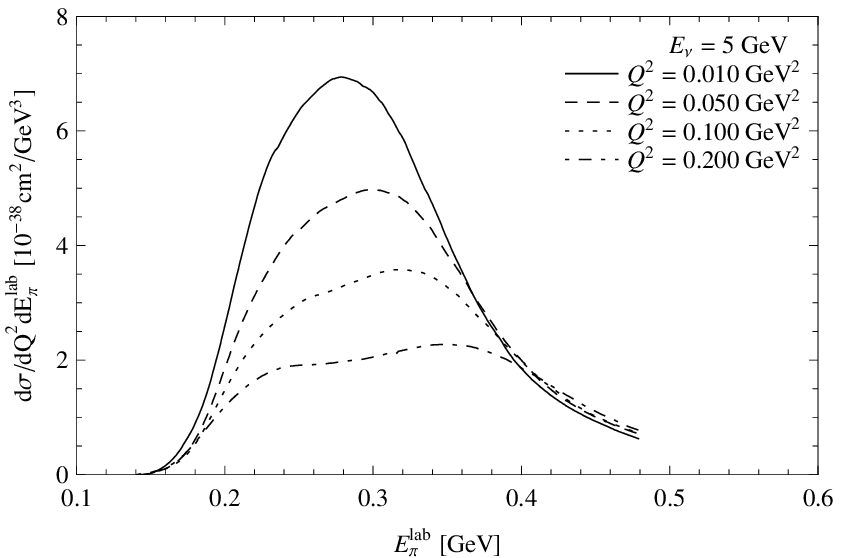}}
\caption{Differential cross section after $W$-integration for various values of momentum transfers $Q^2$}
\label{fig:dsdQsqdEpilabofEpilab}
\end{figure}

\begin{figure}
\subfigure[$E_\nu = 1$ GeV]{\label{fig:dsdQsqdEpilabofQsq1}\includegraphics[width=0.49\textwidth]{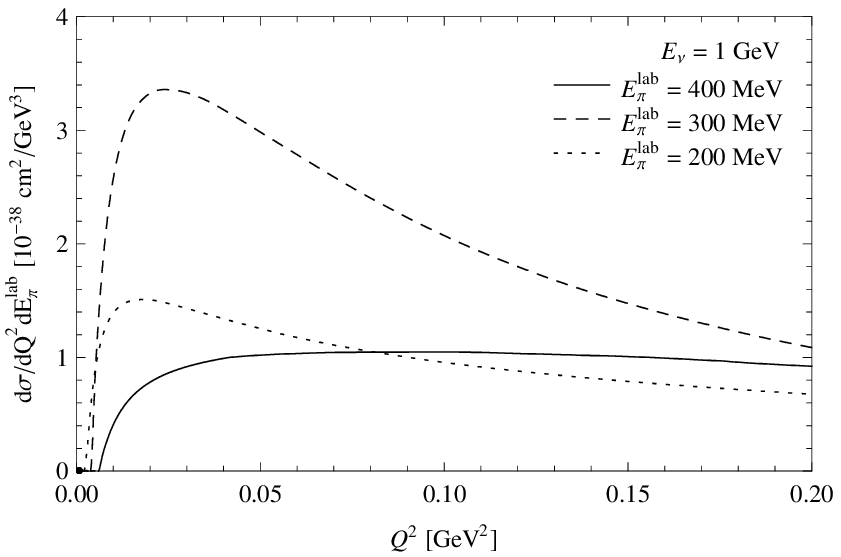}}\hfill
\subfigure[$E_\nu = 5$ GeV]{\label{fig:dsdQsqdEpilabofQsq5}\includegraphics[width=0.49\textwidth]{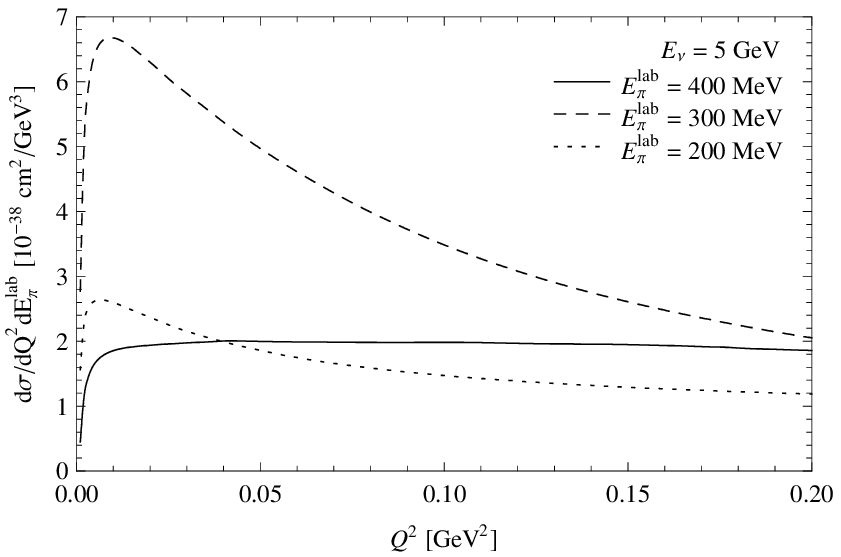}}
\caption{Differential cross section after $W$-integration for fixed pion energies $E_\pi^\textmd{lab}$.}
\label{fig:dsdQsqdEpilabofQsq}
\end{figure}

After integration over $W$ we present the cross section $\frac{\d \sigma^{(A)}}{\d Q^2 \d E_\pi^\textmd{lab}}$ in figure~\ref{fig:dsdQsqdEpilabofEpilab} as a function of $E_\pi^\textmd{lab}$ for four values of $Q^2$. The dependence on $Q^2$ at low energies, shown in figure~\ref{fig:dsdQsqdEpilabofQsq}, is not monotonic but increases up to $Q^2 = \unit[0.050]{GeV^2}$ and then decreases.

\begin{figure}
\subfigure[$E_\nu = 1$ GeV]{\label{fig:dsdQsq1}\includegraphics[width=0.49\textwidth]{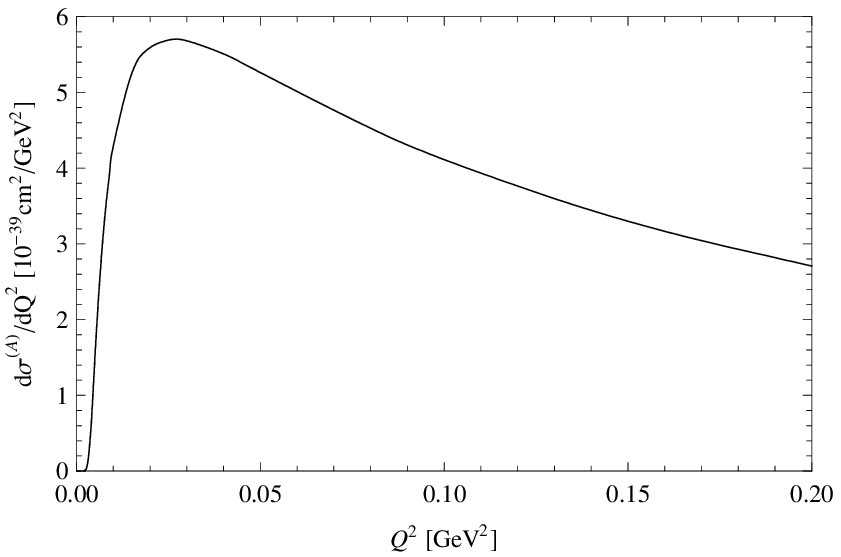}}\hfill
\subfigure[$E_\nu = 5$ GeV]{\label{fig:dsddsdQsq5}\includegraphics[width=0.49\textwidth]{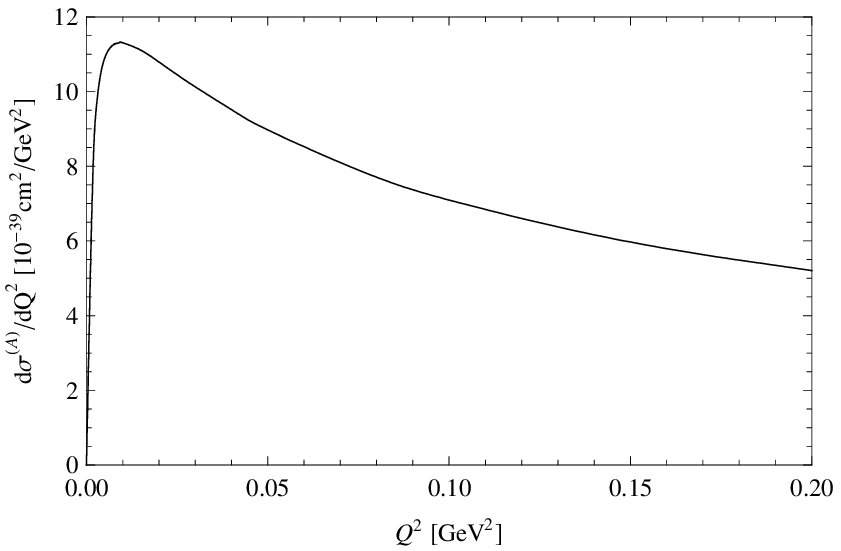}}
\caption{Differential cross section after $W$- and $E_\pi^\textmd{lab}$-integration. Note the cange of scale as $E_\nu$ increases.}
\label{fig:dsdQsq}
\end{figure}

Alternatively, we plot $\frac{\d \sigma^{(A)}}{\d Q^2 \d E_\pi^\textmd{lab}}$ as a function of $Q^2$ for various values of $E_\pi^\textmd{lab}$. In figure~\ref{fig:dsdQsqdEpilabofQsq} we note that the maximal curve is for $E_\pi^\textmd{lab} = \unit[300]{GeV}$ and the peak at $Q^2$ at smaller values than $\unit[0.05]{GeV^2}$. Integrating over $E_\pi^\textmd{lab}$ we obtain the curve in figure~\ref{fig:dsdQsq} with the maximum region being at $Q^2 \approx \unit[0.02]{GeV^2}$, which a posteriori justifies our small $Q^2$ approximation. The single differential cross section $\frac{\d\sigma}{\d Q^2}$ when compared with a recent publication~\cite{Paschos:2011ye} is a little smaller. The reason is that the resonant term $\sigma_{\pi^+p}(W)$ of equation~(\ref{eq:sigmapip}) is computed in this article at $W$, but the data in reference~\cite{Paschos:2011ye} was computed at $\sigma_{\pi^+p}(\nu)$. We repeated the calculation evaluating the data as $\sigma_{\pi^+p}(W)$ and obtained the same result in figure~\ref{fig:dsdQsq} which is our prefered calculation. Three comparisons with experimental data~\cite{AguilarArevalo:2010bm,Radecky:1981fn,Kitagaki:1986ct} were made in our earlier publication~\cite{Paschos:2011ye} and more are expected to be done in the future when new results become available~\cite{Osta,Snider}.

\section{Nuclear Corrections}
\label{sec:NuclearCorrections}
Up to now we presented results for cross sections on free protons and neutrons. The experiments, however,  are carried out on medium size nuclei inside of which the produced resonances decay. Their short propagation in the medium may influence their width~\cite{Marteau:2002sh,Oset:1986sy}. The main modification, however, occurs in the decay products since the pions can be absorbed by the medium or rescatter exchanging their charges some of the time. We shall include these corrections in the Adler-Nussinov-Paschos (ANP) model whose predictions have been tested to some extent qualitatively~\cite{Bolognese:1978yz}.

We describe the process in two steps: the production of the pions and then their propagation through the nucleus. The second step involves a transport matrix that has been calculated in the articles~\cite{Adler:1974qu,Paschos:2007pe}. Both steps satisfy charge symmetry. We discuss first the results of the model and then we shall remark on the results that follow from the cross sections.

We consider isotopically neutral nuclei where the numbers of protons and neutrons are equal. The production of $\pi^+$ proceeds through the standard reaction $\nu_\mu p \rightarrow \mu^- \Delta^{++} \rightarrow \mu^- p \pi^+$ and through
\begin{align}
\sigma ( \nu_\mu n \rightarrow \mu^- \Delta^+ \rightarrow \mu^- n \pi^+ ) = \frac{1}{9} \sigma ( \nu_\mu p \rightarrow \mu^- \Delta^{++} ).
\end{align}
Thus the total yields on protons plus neutrons in terms of the standard cross section are
\begin{align}
\pi^+_i :  & \hspace{5mm} \frac{10}{9}  \sigma ( \nu_\mu p \rightarrow \mu^- \Delta^{++} ) \\
\pi^0_i :  & \hspace{5mm} \frac{2}{9}  \sigma ( \nu_\mu p \rightarrow \mu^- \Delta^{++} ) \\
\pi^-_i :  & \hspace{5mm} \textmd{no direct production}.
\end{align}
We have indicated the pions with a subscript $i$ to indicate that they are produced in the first step of the interaction. These yields must be folded with the charge exhange matrix for C$^{12}$ which has been calculated in the original work~\cite{Adler:1974qu} and also more recently~\cite{Paschos:2007pe}. The pions emerging from the nuclei are indicated with the subscript $f$:
\begin{align}
\begin{pmatrix}
 \pi^+_f \\ \pi^0_f \\ \pi^-_f
\end{pmatrix}_{(p+n)}
& = A
\begin{pmatrix}
0.83	& 0.14 	& 0.04	\\
0.14	& 0.73	& 0.14	\\
0.04	& 0.14	& 0.83	
\end{pmatrix}
\begin{pmatrix}
 \frac{10}{9} \\ \frac{2}{9} \\ 0
\end{pmatrix}
\sigma (\nu_\mu p \rightarrow \mu^- \Delta^{++} ) \\[5mm]
& = A
\begin{pmatrix}
 0.953 \\ 0.318 \\ 0.075
\end{pmatrix}
\sigma (\nu_\mu p \rightarrow \mu^- \Delta^{++} )
\label{eq:ANP}.\end{align}
The overall factor $A$ is for the absorption of the pions and has values from 0.631 for iron to 0.791 for carbon.

A similar analysis follows for antineutrinos with the standard cross section being now $\sigma (\overline{\nu}_\mu n \rightarrow \mu^+ \Delta^- )$. The structure functions of the two standard cross sections are equal through an isospin rotation, however, in the cross section for antineutrinos the vector-axialvector interference has an overall minus sign. For incident antineutrinos the yields are
\begin{align}
\begin{pmatrix}
 \pi^+_f \\ \pi^0_f \\ \pi^-_f
\end{pmatrix}_{(p+n)}^\textmd{anti}
& = A
\begin{pmatrix}
0.83	& 0.14 	& 0.04	\\
0.14	& 0.73	& 0.14	\\
0.04	& 0.14	& 0.83	
\end{pmatrix}
\begin{pmatrix}
 0  \\ \frac{2}{9} \\ \frac{10}{9}
\end{pmatrix}
\sigma (\overline{\nu}_\mu n \rightarrow \mu^+ \Delta^-) \\[5mm]
& = A
\begin{pmatrix}
 0.075 \\ 0.318 \\ 0.953
\end{pmatrix}
\sigma (\overline{\nu}_\mu n \rightarrow \mu^+ \Delta^-)
\label{eq:antiANP}.\end{align}
We derived the results in the model but several propertiees are more general. The difference in the overall cross sections are plotted in figures~\ref{fig:vCCp} and~\ref{fig:avCCn}. As mentioned already the dominant contribution for $\overline{\nu}_\mu n \rightarrow \mu^+ \Delta^-$ is from the axial current, which in the kinematic region of the article was determined by using PCAC.

We notice that the values in the column matrices in equation~(\ref{eq:antiANP}) are the inverted values from those of equation~(\ref{eq:ANP}). This is a consequence of charge symmetry. It follows now that
\begin{align}
\left( \frac{\pi^+}{\pi^0} \right)_\nu = \left( \frac{\pi^-}{\pi^0} \right)_{\overline{\nu}}
.\end{align}
This relation and other ratios are general. They are valid in other kinematic regions outside the PCAC region discussed in this article and they may help the searches for CP asymmetries. Similar results hold for neutral current reactions. We summarize the results for all reactions in table~\ref{tab:nuclear}. The results of nuclear corrections given in the tables can be combined with equations~(\ref{eq:ANP}) and~(\ref{eq:antiANP}) in order to give the final yields for each reaction.

\begin{table}
	\centering
\begin{tabular}{|ccccc|}
  \hline
  	&	\multicolumn{3}{c}{yields} &  \\
  reactions & $\pi^+$	&	$\pi^0$	&	$\pi^-$	& standard cross section	\\
  \hline
  charged currents	& & & & \\
  $\nu_\mu (p+n)$							&	0.953	&	0.318	&	0.075	&	$\sigma (\nu_\mu p \rightarrow \mu^- \Delta^{++})$ \\
  $\overline{\nu}_\mu (p+n)$	&	0.075	&	0.318	&	0.953	&	$\sigma (\overline{\nu}_\mu n \rightarrow \mu^+ \Delta^-)$ \\
  \hline
  neutral currents	& & & & \\
  $\nu_\mu (p+n)$							&	0.473	&	1.06	&	0.473	&	$\sigma (\nu_\mu p \rightarrow \nu_\mu \Delta^+)$ \\
  $\overline{\nu}_\mu (p+n)$	&	0.473	&	1.06	&	0.473	&	$\sigma (\overline{\nu}_\mu p \rightarrow \overline{\nu}_\mu \Delta^+)$ \\
  \hline
\end{tabular}
\caption{Yields for neutrinos and antineutrinos including nuclear corrections.}
\label{tab:nuclear}
\end{table}

\section{Anaysis of the yields}
\label{sec:AnaysisOfTheYields}
The results of this study point to general properties which may be useful to other investigations. For the determination of oscillation parameters and the observation of CP asymmetries we need the yields of single pions from nuclear targets which have the same energy dependence for neutrino and antineutrino reactions. We point out that for isoscalar targets the nuclear corrections for $\pi^+$ and $\pi^-$ are equal provided that their energy spectrum at the production vertex are the same. This requires the production by neutrinos and antineutrinos to be the same.

\begin{figure}
\subfigure[$\nu_\ell p \rightarrow \ell^- \Delta^{++}$]{\label{fig:vCCpELECTRON}\includegraphics[width=0.49\textwidth]{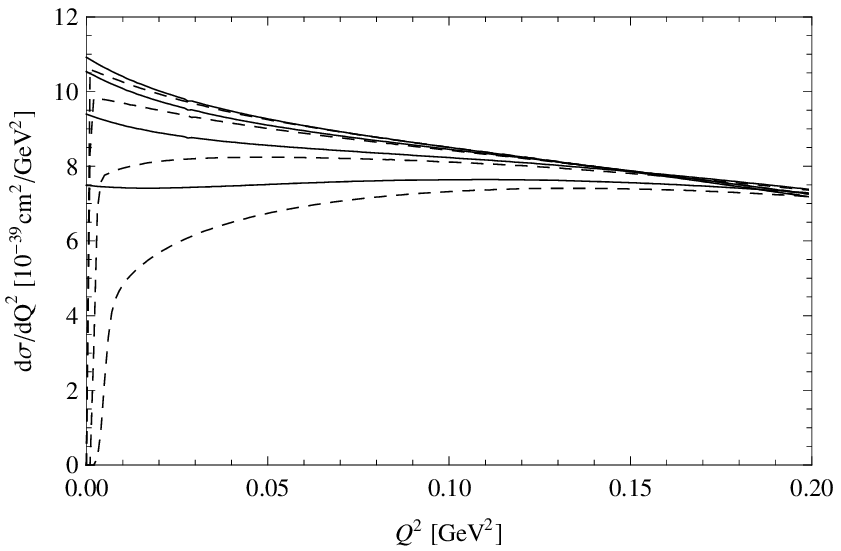}}\hfill
\subfigure[$\overline{\nu}_\ell n \rightarrow \ell^+ \Delta^-$]{\label{fig:avCCnELECTRON}\includegraphics[width=0.49\textwidth]{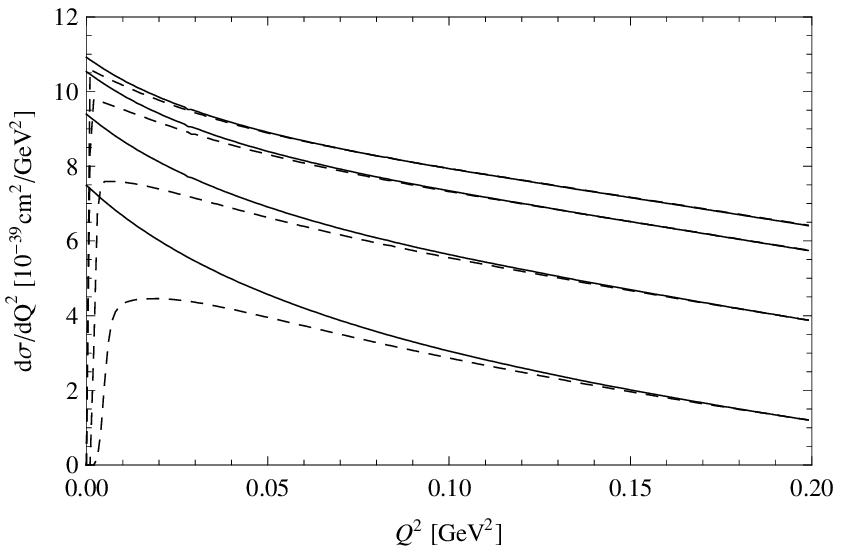}}\hfill
\caption{Comparison of muon and electron neutrino or antineutrino cross section for various energies $E_{\nu,\overline{\nu}}=\unit[1, 2, 5, 10]{GeV}$ (bottom to top). The solid line is for electron neutrinos or antineutrinos and the dashed one for muon neutrinos and antineutrinos, respectively.}
\label{fig:electron}
\end{figure}

We pointed out that the structure functions for reactions~(\ref{eq:vCCp}) and~(\ref{eq:avCC}) are the same by charge symmetry. The difference in the cross sections comes from the sign in front of $\mathcal{W}_3$. We remark now that the contribution of $\mathcal{W}_3$ to the cross sections diminishes as $E_{\nu,\overline{\nu}}$ increases. This is evident in equation~(2.10) of reference~\cite{Lalakulich:2005cs} and is apparent in figures~6~--~8 of reference~\cite{Paschos:2008gs}. The merging together of neutrino and antineutrino cross sections also occurs as $Q^2\rightarrow 0$. We pointed this out for neutral currents in figure~\ref{fig:IsospinRelationsNC}. It also happens in charged current reactions and we demonstrate it for $\nu_e p \rightarrow e^- \Delta^{++}$ and $\overline{\nu}_e n \rightarrow e^+ \Delta^-$ by repeating the calculation and replacing the muon mass by the electron mass. In figure~\ref{fig:electron} we show the two cross sections together and for various energies. For the subsequent interaction the nuclear corrections on isoscalar targets will be the same for $\pi^+$ and $\pi^-$.

These remarks suggest to use several reactions in the nearby detector in order to determine the fluxes and at the same time check the shapes and normalizations of pion production. Then measure in the far away detector charge-symmetric reactions. Ratios that follow from charge symmetry should provide checks at every step. For instance, in the charged current reactions the $\pi^+$ yield is dominant, but for antineutrinos the $\pi^-$ yield dominates. For neutral currents the $\pi^0$ yield is dominant and the sum $\pi^++\pi^-$ is, within 5~\%, equal to the $\pi^0$ yield.

Then one can use the determined fluxes of $\nu_\mu$ and $\overline{\nu}_\mu$ in order to compute the regenerated, through oscillations, $\nu_e$ and $\overline{\nu}_e$ fluxes far away.

For charge symmetric reactions for which the products cross sections $\otimes$ nuclear corrections are the same, the change in the yields at the far away detector should be proportional to the change in the fluxes, coming from the oscillations with or without CP violation.

\section{Discussion}
\label{sec:Discussion}
We carried out an extensive program of calculations in the kinematic region $m_N\nu \gsim Q^2$ and $Q^2\leq \unit[0.20]{GeV^2}$, where PCAC gives the dominant contribution for the axial current. We give results for charged and neutral current reactions of neutrinos and antineutrinos. One interesting result, in the energy range $E_\nu = \unit[1 - 2]{GeV}$, is that the axial contribution dominates the $\overline{\nu}_\mu n \rightarrow \mu^+ \Delta^-$ reaction. In the calculations we kept the mass $m_\mu$ which produces the turn over in the differential cross sections $\frac{\d\sigma}{\d Q^2}$ as $Q^2\rightarrow 0$. The long baseline oscillation experiments will be searching for $\nu_e p \rightarrow e^- \Delta^{++}$ type reactions where $m_e \ll m_\mu$ and we repeated the results in figure~\ref{fig:electron}. The PCAC limiting point at $Q^2=0$ is now finite.

The energy spectra for the pions are also interesting and measurable. In section~\ref{sec:EnergySpectrumAndAngularDependenceOfPions} we outlined a method and carried out the calculations obtaining the energy and angular spectra of the pions with the help of transformations summarized in the appendix. Since our estimates for the axial current are valid for $Q^2\lsim \unit[0.20]{GeV^2}$ we limited the figures to that kinematic region.

For nuclear corrections we adopted the ANP model which is both simple and transparent. We presented corrections for the incoherent sum of a proton and neutron target. They can be scaled up for other isoscalar nuclei (C$^{12}$, O$^{16}$, ...) by multiplying with the appropriate number of particles in the target and using the appropriate transfer matrix. Then we pointed out that some general properties follow from charge symmetry. These relations follow from isospin rotations and should hold beyond the kinematic domains introduced by the validity of PCAC.

We hope that the method and results can be adopted in oscillation experiments searching for CP asymmetry in the leptonic sector.

\section*{Acknowledgements}
\label{sec:Acknowledgements}
One of us (EAP) wishes to thank Drs. W. Bardeen and S. Parke for the hospitality at Fermi Laboratory and the Humboldt Foundation for a traveling grant. Numerous discussions with Dr. J. Morfin and members of the MINERvA group helped us formulate the physical issues of this work.

\section*{Appendix: Lorentz transformation from cm to lab frame}
\label{sec:LorentzTransformationFromCmToLabFrame}
We define as center of mass system the frame where the resonance is at rest. In the laboratory the four-momentum of the excited resonance after the collision is 
\begin{align}
p_\mu' = ( \nu + m_N, |\vec{q}| ). 
\end{align}
We can bring the resonance to rest by the transformation parameter
\begin{align}
\beta = \frac{|\vec{q}|}{\nu + m_N}
\end{align}
and the corresponding
\begin{align}
\gamma = \frac{1}{\sqrt{1-\beta^2}} = \frac{\nu + m_N}{W}.
\end{align}
In the rest frame of the resonance with its invariant mass within its width by $W$, the energy of the pion is
\begin{align}
E_\pi^\textmd{cm} = \frac{W^2-m_N^2+m_\pi^2}{2W}.
\end{align}
With this information we can relate various quantities in the two frames.

The scattering angles of the pion satisfy
\begin{align}
\tan \theta_\textmd{lab} = \frac{\sin \theta_\textmd{cm}}{\gamma\left( \cos \theta_\textmd{cm} + \gamma \frac{E_\pi^\textmd{cm}}{p_\pi^\textmd{cm}} \right)}.
\end{align}
The differential cross sections are related by
\begin{align}
\frac{\d \sigma}{\d \Omega_\textmd{lab}} = \frac{\d \sigma}{\d \Omega_\textmd{cm}} \frac{1}{J}
\label{eq:apa}\end{align}
and
\begin{align}
\frac{\d \sigma}{\d E_\pi^\textmd{lab}} = 2 \pi \frac{\d \sigma}{\d \Omega_\textmd{lab}} \left[ \frac{(m_N+\nu)p_\pi^\textmd{lab} - |\vec{q}| E_\pi^\textmd{lab} \cos \theta_\textmd{lab}}{|\vec{q}| (p_\pi^\textmd{lab})^2} \right]
\label{eq:apb}\end{align}
with the Jacobian $J$ derived from the above equation
\begin{align}
J = \frac{\d \cos \theta_\textmd{lab}}{\d \cos \theta_\textmd{cm}} = \frac{\gamma \left( 1 + \beta \frac{E_\pi^\textmd{cm}}{p_\pi^\textmd{cm}} \cos \theta_\textmd{cm} \right)}{\left( \sin^2 \theta_\textmd{cm} + \gamma^2 \left( \cos \theta_\textmd{cm} + \frac{\beta E_\pi^\textmd{cm}}{p_\pi^\textmd{cm}} \right)^2 \right)^{\nicefrac{3}{2}}}.
\label{eq:apc}\end{align}
Finally, we obtain
\begin{align}
\frac{\d \cos \theta_\textmd{lab}}{\d E_\pi^\textmd{lab}} = \frac{(m_N+\nu) p_\pi^\textmd{lab} - |\vec{q}| E_\pi^\textmd{lab} \cos \theta_\textmd{lab}}{|\vec{q}| (p_\pi^\textmd{lab})^2}.
\end{align}

\end{document}